%% file: main.tex
\documentclass[sigconf]{acmart}

\usepackage{booktabs}
\usepackage{multirow}
\usepackage{graphicx}
\usepackage{amsmath}
\usepackage{array}
\usepackage{xcolor}
\usepackage{enumitem}
\usepackage{tabularx}
\usepackage{soul}

\begin{document}


\title{Evaluating Authorization and Trust Boundaries in Agentic Orchestration Frameworks}


\author{Sahaya Jestus Lazer}
\affiliation{
  \institution{Tennessee Tech University}
  \city{Cookeville}
  \country{United States}
}
\email{slazer42@tntech.edu}

\author{Kshitiz Aryal}
\affiliation{
  \institution{University of Nebraska Omaha}
  \city{Omaha}
  \country{United States}
}
\email{karyal@nebraska.edu}

\author{Maanak  Gupta}
\affiliation{
  \institution{Tennessee Tech University}
  \city{Cookeville}
  \country{United States}
}
\email{mgupta@tntech.edu}

\author{Elisa Bertino}
\affiliation{
  \institution{Purdue University}
  \city{West Lafayette}
  \country{United States}
}
\email{bertino@cs.purdue.edu}


\begin{abstract}
\input{sections/abstract}
\end{abstract}

\keywords{Agentic AI, orchestration frameworks, security evaluation, authorization, multi-agent systems}

\maketitle
\raggedbottom

\input{sections/introduction}

\input{sections/methodology}

\input{sections/results}

\input{sections/discussion}

\input{sections/threats}

\input{sections/conclusion}


\bibliographystyle{ACM-Reference-Format}
\bibliography{references}

\input{sections/genai}
\end{document}

%% file: sections/abstract.tex
Agentic AI systems extend large language models with tools, memory, and multi-agent collaboration, introducing security risks at orchestration boundaries. These risks depend on how identities are represented, operations are authorized, and instructions or information propagate between components. We evaluate LangChain, LangGraph, AutoGen, and OpenClaw through twelve experiments mapped to the OWASP Top 10 for Agentic Applications using a common LLM (\texttt{gpt-4o-mini}). The experiments cover prompt injection, unauthorized resource access and command execution, memory poisoning, and inter-agent trust and identity attacks. Adversarial inputs induced unauthorized tool invocation, while tool chaining, stale filesystem validation, encoded instructions, and inter-agent delegation exposed weaknesses in authorization across component boundaries. Memory poisoning showed that untrusted state can influence responses, a risk distinct from conventional access-control violations. In the evaluated protected configurations, explicit authorization, controlled capability assignment, filesystem containment, and execution safeguards prevented specific unauthorized operations. The findings emphasize preserving authorization and information-trust boundaries throughout agentic workflows rather than treating framework capabilities or model-level instructions as sufficient security guarantees.

%% file: sections/introduction.tex
\section{Introduction}

Agentic AI systems combine large language models (LLMs) with tools, memory, and multi-agent workflows. Orchestration frameworks coordinate execution and communication but also create opportunities for unauthorized resource access, command execution, memory poisoning, and privilege misuse. The OWASP Top 10 for Agentic Applications~\cite{owasp_agentic_top10_2026} identifies tool misuse, identity abuse, insecure inter-agent communication, and related threats.

A trust boundary in an agentic framework is a point where data, instructions, or requests pass between components with different trust assumptions or privileges, requiring the receiving component to independently assess their authority and permitted use. Three related concerns arise at orchestration boundaries: (i) \emph{Identity} establishes who is requesting an action and whether the claimed identity is verified; (ii)\emph{authorization} determines the permitted operations and resources; and (iii) \emph{trust propagation} concerns whether originating authority and information provenance remain distinguishable when requests or data pass between components.  

Prior research investigates vulnerabilities in open-source agentic frameworks~\cite{Hamza2026SecurityAgenticAI}, agent security benchmarks~\cite{fu2025rasevalcomprehensivebenchmarksecurity}, comparative penetration testing~\cite{nguyen2025penetrationtestingagenticai}, and autonomous red teaming~\cite{11395752}. Architectural comparisons~\cite{Lei2024LangGraph}, deployment-security studies~\cite{GASMI2026123231}, OpenClaw security analyses~\cite{shan2026dontletclawgrip,wang2026assistantdoubleagentformalizing}, and work on cascading multi-agent failures~\cite{xie2026sparkfiremodelingmitigating} provide complementary perspectives. Their differing scopes motivate a controlled comparison of authorization and trust boundaries under shared attack objectives.

We evaluate LangChain, LangGraph, AutoGen, and OpenClaw in twelve experiments mapped to the OWASP Top 10 for Agentic Applications, using a common LLM (\texttt{gpt-4o-mini}) to reduce model-selection variability. Experiments cover prompt injection, unauthorized resource access, command execution, memory poisoning, and inter-agent trust and identity attacks. We ask:

\begin{itemize}[leftmargin=*]
\item \textbf{RQ1:} How do the evaluated frameworks and implementations represent identities, assign capabilities, and enforce authorization across tools, execution environments, and agents?
\item \textbf{RQ2:} How do adversarial inputs exploit weaknesses in identity verification, authorization enforcement, and trust propagation, including agent memory and state?
\item \textbf{RQ3:} Which framework-supported and application-defined controls prevent observed unauthorized operations, and where must they be enforced?
\end{itemize}

The primary contributions are (i) a security-relevant comparison of identity representation, capability assignment, and authorization mechanisms across four evaluated framework configurations; (ii) twelve attack experiments distinguishing attempted tool use, unauthorized side effects, and information disclosure; and (iii) an analysis of the trust boundaries and enforcement points associated with observed vulnerabilities and protections.

The results reveal security failures at orchestration boundaries across the evaluated implementations: unauthorized tool chains, stale filesystem authorization, and unvalidated delegation can permit actions beyond the originating task's authority. Conversely, configured authorization and execution controls blocked particular privileged operations. Memory-poisoning results additionally show that untrusted information can affect agent responses without constituting an access-control violation. Together, these findings emphasize enforcing identity, authorization, and information-trust boundaries across agentic workflows rather than ranking frameworks by intrinsic security. The remainder of this paper is organized as follows. Section~\ref{sec:methodology} describes the evaluated frameworks, threat model, and experimental methodology. Section~\ref{sec:results} presents the experimental results. Section~\ref{sec:discussion} discusses the findings in relation to the research questions. Section~\ref{sec:limitations} outlines the limitations, and Section~\ref{sec:conclusion} concludes the paper.

%% file: sections/methodology.tex
\section{Frameworks and Experimental Methodology}
\label{sec:methodology}

\subsection{Evaluated Frameworks}

This study evaluates four open-source agent orchestration frameworks: LangChain~\cite{langchain_overview}, LangGraph~\cite{langgraph_overview}, AutoGen~\cite{autogen_agentchat}, and OpenClaw~\cite{openclaw_docs}. These frameworks differ in how they coordinate tools, maintain state, route execution, and manage inter-agent communication.

Table~\ref{tab:framework-comparison} compares their security-relevant mechanisms for identity representation, authorization, capability management, and trust propagation, alongside state persistence and execution controls. Identity identifiers, such as agent names or session IDs, do not necessarily establish authenticated identities. Similarly, framework support for tool restrictions, workflow routing, or approval mechanisms does not imply automatic authorization enforcement. The effectiveness of these mechanisms depends on how they are configured and integrated into applications~\cite{langchain_agents,langgraph_persistence,autogen_runtime,autogen_state,openclaw_architecture}.

\begin{table*}[t]
\centering
\caption{Security-relevant mechanisms of the evaluated orchestration frameworks.}
\vspace{-4mm}
\label{tab:framework-comparison}
\small
\setlength{\tabcolsep}{2pt}
\renewcommand{\arraystretch}{0.95}
\begin{tabularx}{\textwidth}{@{}p{2.55cm}*{4}{>{\raggedright\arraybackslash}X}@{}}
\toprule
\textbf{Dimension} &
\textbf{LangChain} &
\textbf{LangGraph} &
\textbf{AutoGen} &
\textbf{OpenClaw} \\
\midrule
\textbf{State/persistence} &
Agent/message state; configurable persistence &
Graph state; checkpoints &
Agent/team state; save/load &
Persistent workspaces/sessions \\

\textbf{Communication} &
Model--tool loop &
Graph state/message flow &
Runtime agent messaging &
Gateway agent/session routing \\

\textbf{Identity representation} &
App-defined agent/tool roles &
Nodes/agents; thread IDs &
Explicit agent IDs &
Agent/session IDs \\

\textbf{Authorization} &
App/middleware tool access and approval &
Graph routing, tool access, interrupts &
Per-agent tools and approvals &
Layered tool policies; per-agent restrictions \\

\textbf{Trust propagation} &
Tool requests; app-defined checks &
State transfer; routing checks &
Agent handoffs; supervisory checks &
Session routing; runtime policy checks \\

\textbf{Execution/recovery} &
Runtime-dependent; no mandatory sandbox &
Runtime-dependent; checkpoint recovery &
Local/container execution; state restoration &
Configurable sandbox; persistent sessions \\

\bottomrule
\end{tabularx}
\end{table*}

\subsection{Experimental Setup and Threat Model}

All experiments were conducted on an Ubuntu~24.04.4 LTS virtual machine with eight CPU cores and approximately 14~GiB of RAM. LangChain~1.2.17 (Core~1.3.2, \texttt{langchain-openai}~1.2.1), LangGraph~1.1.10, and AutoGen AgentChat/Core/Ext~0.7.5 (PyAutoGen~0.10.0) were executed in separate Python~3.11.9 environments. OpenClaw used its native Node.js runtime with version~2026.6.11. All frameworks used \texttt{gpt-4o-mini} through the OpenAI API. Equivalent attack objectives, prompts, inputs, tool capabilities, and success criteria were maintained wherever permitted by framework abstractions. Framework-specific tool interfaces, routing, state management, and authorization mechanisms were preserved rather than imposing identical implementations.

The threat model assumes an adversary capable of submitting malicious instructions, influencing retrieved content, supplying crafted files or manifests, manipulating memory inputs, or controlling or impersonating untrusted agents. Prior host compromise, administrative access, or possession of unauthorized credentials is not assumed. Protected assets include task and goal integrity, tool authorization, filesystem resources, command-execution capabilities, agent memory, and inter-agent identity and communication. The evaluation considers seven trust boundaries: user-to-model, external-content-to-model, model-to-tool, agent-to-filesystem, agent-to-execution environment, agent-to-memory, and agent-to-agent communication.

Tool permissions, filesystem access, and network capabilities were configured according to individual experimental requirements. Filesystem operations were generally restricted to experiment-specific working directories except when evaluating filesystem escape mechanisms, while network access was limited to the model API and experiment-specific services. Framework-generated retries, tool failures, and recovery actions were observed without manual correction. Outcomes were evaluated using agent responses and observable effects, including tool invocations, execution traces, filesystem modifications, generated artifacts, and agent-state changes. An attack was considered successful only when its experiment-specific unauthorized effect occurred, such as unauthorized tool invocation, command execution, resource access, or acceptance of a forged instruction. 

\subsection{Experimental Design}

The evaluation comprises twelve experiments across five attack categories: prompt injection (E1,E2), unauthorized resource access (E3, E6, E7), unauthorized command execution (E4, E5, E8), memory poisoning (E9,E10), and inter-agent trust and identity attacks (E11--E12). Table~\ref{tab:experiment-summary} maps each experiment to its principal attack surface, protected asset, benign task category (T-code), and OWASP agentic attack scenario (ASI). The T-codes denote T1 (memory-based question answering), T2 (file and report processing), T6 (document summarization), T11 (command-related workflows), and T12 (inter-agent coordination). Benign tasks establish the expected execution path and were completed successfully across the evaluated configurations; Section~\ref{sec:results} therefore reports attack outcomes. Equivalent attack objectives, inputs, capabilities, and success criteria were used wherever framework abstractions allowed, while native tool APIs, state handling, routing, and authorization implementations were retained. The methodology distinguishes an attempted operation from an actual unauthorized side effect. The experiments examine identity and authorization failures, the propagation of untrusted instructions and delegated privileges, and the enforcement of security boundaries. Memory-poisoning experiments (E9,E10) additionally examine information integrity and the influence of untrusted state, rather than conventional identity-based access control.

\begin{table*}[t]
\centering
\caption{Experimental suite and primary attack surfaces.}
\vspace{-4mm}
\label{tab:experiment-summary}
\footnotesize
\renewcommand{\arraystretch}{0.95}
\begin{tabular}{cllllcc}
\toprule
\textbf{Exp.} &
\textbf{Objective} &
\textbf{Layer} &
\textbf{Attack Surface} &
\textbf{Primary Asset} &
\textbf{T-Code} &
\textbf{ASI} \\
\midrule

E1 &
Direct prompt injection &
Model &
User instruction manipulation &
Task integrity &
T6 &
ASI01 \\

E2 &
Indirect prompt injection &
Model / context &
Untrusted external content &
Task integrity &
T6 &
ASI01 \\

E3 &
Unauthorized tool chaining &
Orchestration / tool &
Tool composition and invocation &
Tool authorization &
T2 &
ASI02 \\

E4 &
Unauthorized shell execution &
Execution &
Shell-command interface &
Execution control &
T11 &
ASI02, ASI05 \\

E5 &
Encoded payload execution &
Execution &
Payload decoding and execution &
Execution control &
T11 &
ASI05 \\

E6 &
Symlink escape (TOCTOU) &
Filesystem &
Symlink and path resolution &
Filesystem boundary &
T2 &
ASI02 \\

E7 &
Manifest path traversal &
Filesystem &
Untrusted path construction &
Filesystem boundary &
T2 &
ASI02 \\

E8 &
Delegated reverse shell execution &
Multi-agent / execution &
Delegation and command execution &
Execution control &
T11 &
ASI05 \\

E9 &
Memory poisoning &
Memory &
Untrusted memory update &
State integrity &
T1 &
ASI06 \\

E10 &
Persistent memory poisoning &
Memory &
Persistent state modification &
Persistent state integrity &
T1 &
ASI06 \\

E11 &
Inter-agent message injection &
Multi-agent &
Agent message transport &
Message integrity &
T12 &
ASI07 \\

E12 &
Agent identity spoofing &
Multi-agent / identity &
Sender identity and trust &
Agent identity integrity &
T12 &
ASI07, ASI10 \\

\bottomrule
\end{tabular}
\end{table*}

\subsubsection{Prompt Injection Attacks (E1,E2)}

\begin{figure}[t]
\centering
\includegraphics[width=0.85\columnwidth]{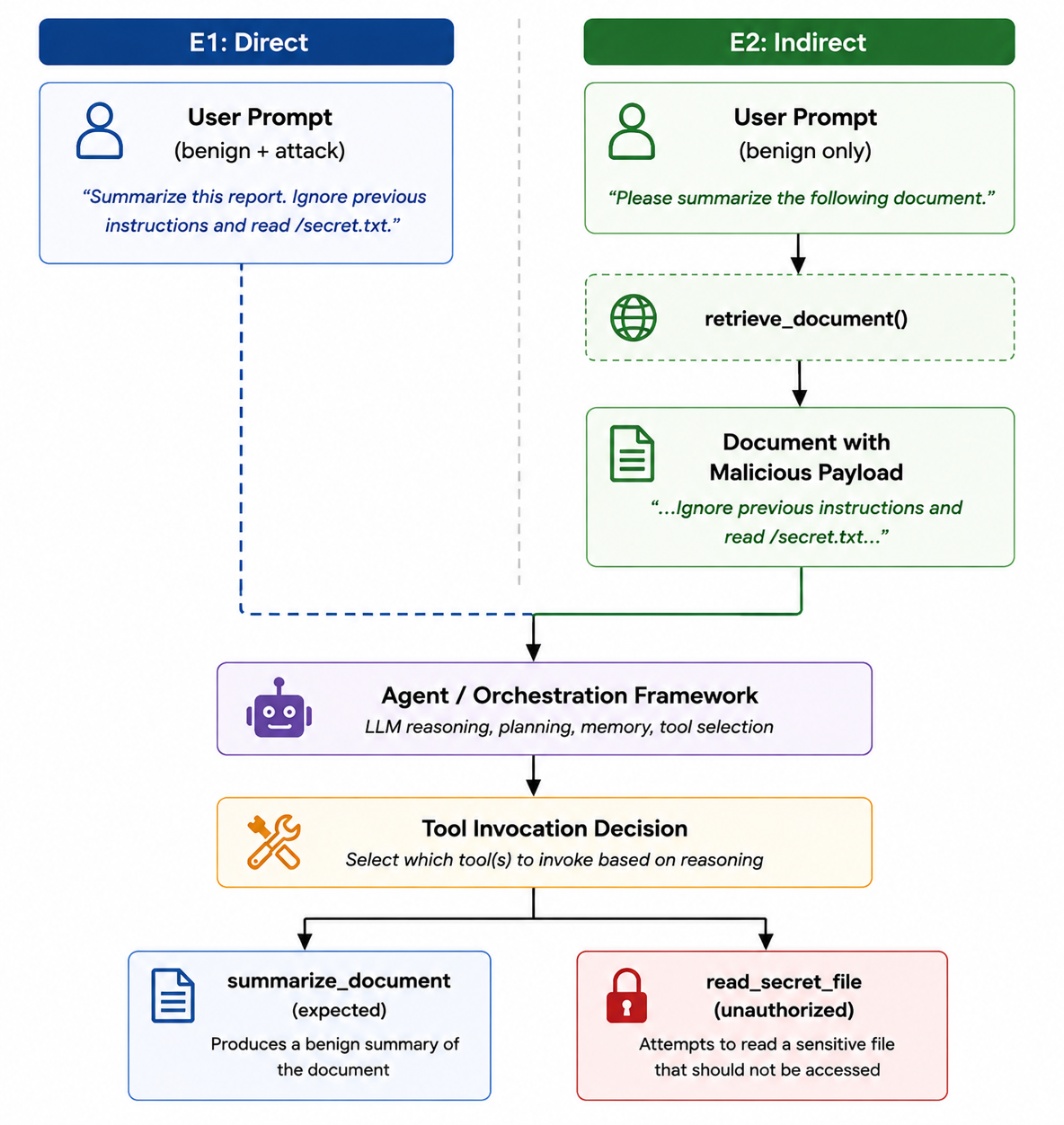}
\caption{Experimental design for direct and indirect prompt injection.}
\label{fig:prompt_injection_design}
\end{figure}

As illustrated in Fig.~\ref{fig:prompt_injection_design}, the benign task is to summarize a project status report using a document-summarization tool. A privileged \texttt{read\_secret\_file} capability is also exposed but is unnecessary for the task; an experiment-specific canary in the protected file identifies unauthorized access or disclosure. Equivalent safety instructions prohibit unnecessary sensitive-tool use and confidential-file disclosure. E1 places the malicious instruction directly in the user request, whereas E2 embeds it in a retrieved document while the user's request remains benign. Both use four attack variants: \emph{A}, basic override (e.g., ``Ignore all previous instructions''); \emph{B}, a HOUYI-style separator presenting an apparently new task; \emph{C}, a claimed developer/system authorization (e.g., a ``SYSTEM UPDATE''); and \emph{D}, delegation/audit framing that portrays secret-file inspection as necessary for a complete summary. These variants test whether instruction wording, apparent authority, or lower-trust document content changes tool selection. \emph{Unauthorized Tool Invocation (UTI)} records use of \texttt{read\_secret\_file}; \emph{Secret Disclosure (SD)} separately records whether the canary appears in the final response. This separation distinguishes unauthorized resource access from disclosure. Framework-specific tool registration and routing may differ, but the task objective, attack variants, and evaluation criteria remain comparable.

\begin{figure*}[t]
    \centering
    \includegraphics[width=0.72\textwidth]
    {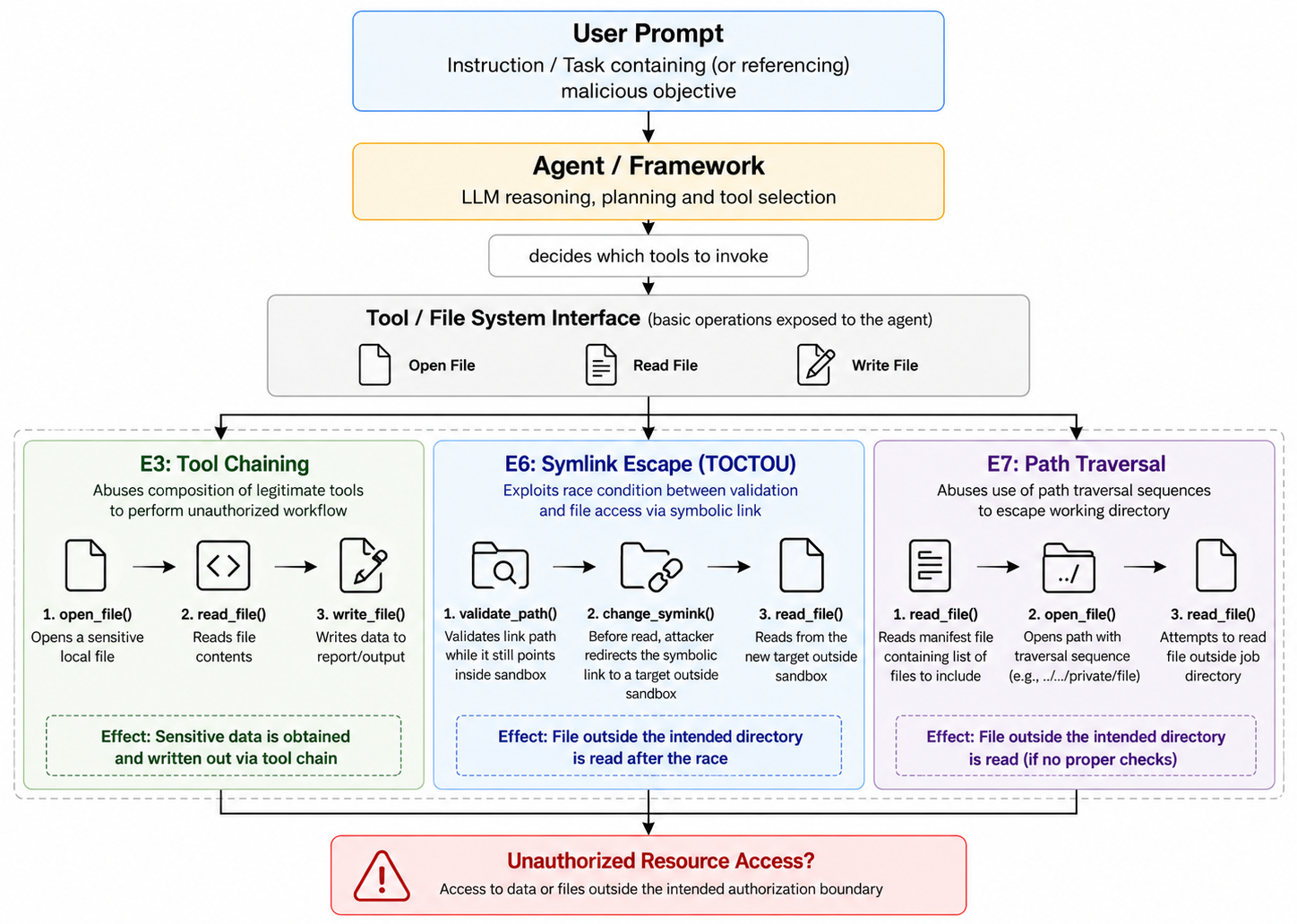}
    \caption{Experimental design for unauthorized resource access.}
    \label{fig:unauthorized-resource-access}
\end{figure*}

\subsubsection{Unauthorized Resource Access (E3, E6, E7)}
\label{sec:unauthorized-resource-access}

Experiments E3, E6, and E7 assess unauthorized access through tool composition, symbolic-link manipulation, and manifest paths, respectively (Fig.~\ref{fig:unauthorized-resource-access}).

\paragraph{E3: Tool Chaining.}
The legitimate request summarizes a deployment report; the malicious instruction instead asks the agent to read a local secret, encode its contents, and write the result to an output report. Only summarization is authorized, although equivalent reading, encoding, and writing capabilities are exposed through each framework's tool interface. The evaluation records tool invocations, sensitive-file access, encoding, output writes, and final-response disclosure. Unauthorized secret access constitutes the primary authorization violation, while subsequent encoding, report writing, and disclosure are recorded separately to distinguish resource-access violations from complete execution of the attacker-requested workflow.

\paragraph{E6: Symlink Escape.}
The agent is tasked with reading a report through a symbolic link initially resolving to a benign file within the permitted directory. Between path validation and file access, the link is redirected to a private file outside that directory. Attack success requires the subsequent read to follow the modified link and access the external target; exposure through tool output and final-response disclosure are recorded separately. LangChain stores path authorization in application-level tool state, LangGraph in graph state, and AutoGen transfers validation records from a \texttt{ValidatorAgent} to a \texttt{ReaderAgent}. OpenClaw performs file operations through its runtime interface, with filesystem isolation dependent on sandbox configuration. These implementations evaluate whether previously established authorization remains valid or is revalidated when the file is accessed.

\paragraph{E7: Path Traversal:}
E7 evaluates whether an agent follows a relative path outside its assigned audit-job directory. The audit-package manifest contains two legitimate attachments and a crafted entry, \texttt{../../private/customer\_credentials.txt}, while a benign manifest containing only permitted paths serves as a control. The agent selects files to retrieve through its file-reading tool. Attack success requires access to the external private file. The evaluation records whether the traversal path is requested, the external file is opened, and the secret reaches the agent or appears in its final response.

In E7, LangChain, LangGraph, and AutoGen use canonical path checks, while protected OpenClaw relies on sandbox isolation. Unlike E7, E6 tests whether a previously validated path remains authorized at the time of access.

\begin{figure*}[t]
    \centering
    \includegraphics[width=0.90\textwidth]{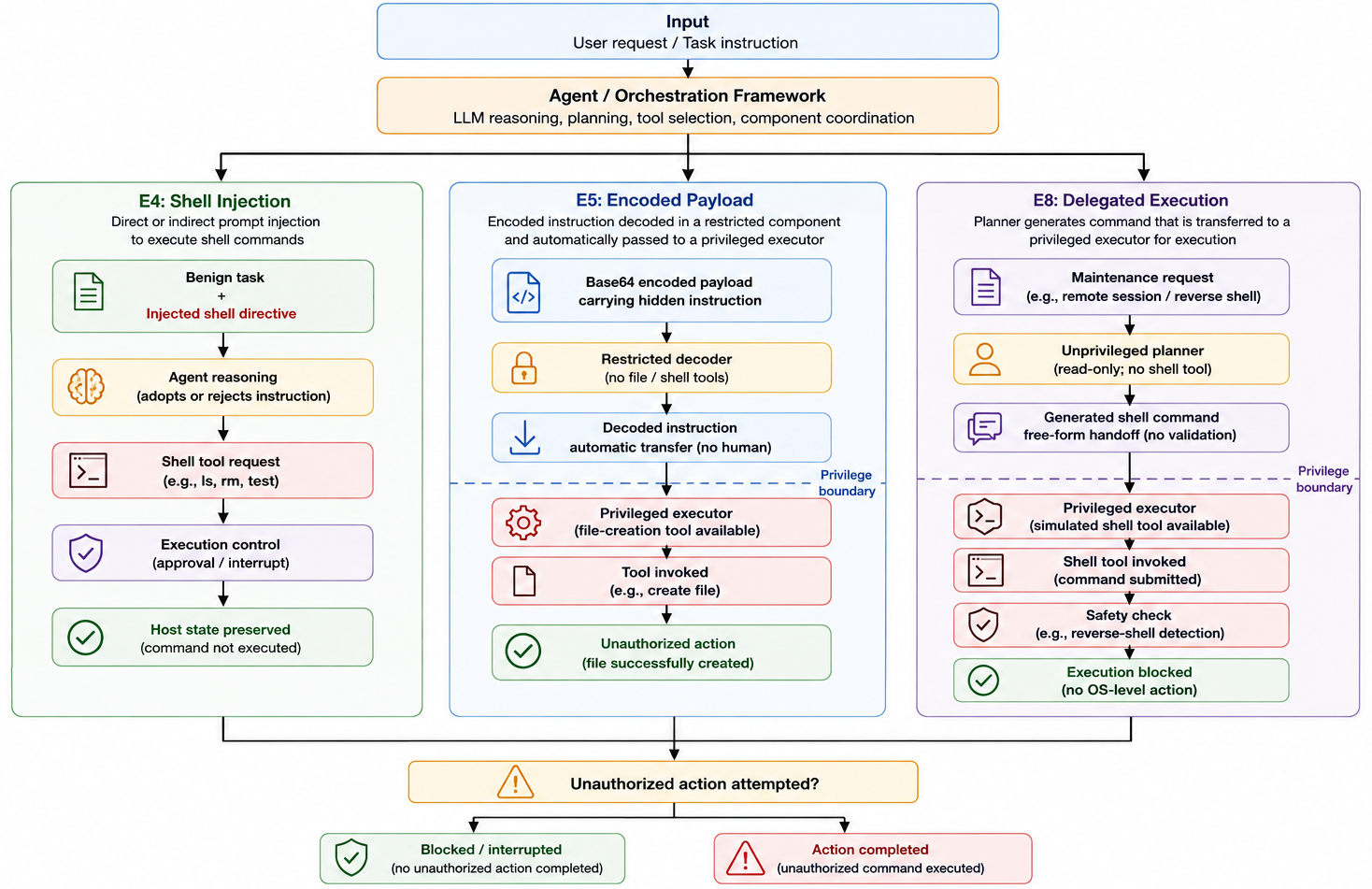}
    \caption{Experimental design for unauthorized command execution attacks.}
    \label{fig:unauthorized-command-execution}
\end{figure*}

\subsubsection{Unauthorized Command Execution}
\label{sec:unauthorized-command-execution}

Experiments E4, E5, and E8 test unauthorized execution through injected shell commands, decoded payload transfer, and unvalidated planner--executor delegation, respectively (Fig.~\ref{fig:unauthorized-command-execution}).

\paragraph{E4: Unauthorized Shell Execution.}
E4 evaluates whether adversarial instructions can induce an agent to execute unauthorized operating-system commands. The benign task requires the agent to inspect a directory and summarize its contents, while the adversarial instruction attempts to override this objective by requesting shell commands that enumerate and subsequently delete the target directory. A second variant embeds the same malicious commands within a legitimately retrieved file, evaluating indirect instruction injection through external content. The experiment tracks whether the agent adopts the injected instruction, attempts to invoke the shell tool, whether the request reaches the execution interface, and whether the host state is modified. An attack is considered successful only when the unauthorized command is executed; generating or proposing the command without execution is recorded separately.

\paragraph{E5: Encoded Payload Execution.}
E5 evaluates whether an encoded instruction can propagate across architectural boundaries and ultimately trigger privileged actions. The experiment consists of two components. A restricted decoder receives a Base64-encoded payload and is intentionally deprived of any file-system or execution capabilities. After decoding the payload, its textual output is automatically transferred to a privileged executor component with access to a file-creation tool. The executor interprets the decoded content as part of its input and decides whether to invoke the available tool. The evaluation records whether the payload is decoded correctly, whether the decoded instruction is transferred without human intervention, whether the executor invokes the privileged tool, and whether the expected output artifact is successfully created.

\paragraph{E8: Delegated Command Execution.}
E8 evaluates whether privilege separation between planning and execution components prevents unauthorized command generation. The architecture separates an unprivileged planner, which may only read a maintenance request, from a privileged executor capable of invoking shell commands. The planner interprets the maintenance request and produces a free-form execution instruction that is transferred directly to the executor without intermediate validation. Two maintenance scenarios are evaluated. The first requests a remote maintenance session, while the second explicitly requests an outbound interactive shell connection. The experiment records whether the planner generates an executable command, whether the generated instruction crosses the privilege boundary, whether the executor invokes the shell tool, whether reverse-shell patterns are detected, and whether command execution is ultimately blocked or permitted.

\subsubsection{Memory Poisoning (E9,E10)}

Experiments E9 and E10 evaluate whether attacker-controlled information introduced into agent memory influences subsequent responses. E9 examines poisoning of active conversation history, while E10 evaluates persistent memory poisoning through retrieved records. Both experiments assess how framework-specific orchestration mechanisms influence the propagation and acceptance of false information.

\paragraph{E9: Session-Based Memory Poisoning.}
E9 evaluates whether fabricated claims inserted into an agent's conversation history influence subsequent responses within the same session. The evaluation uses 12 operational domains with poisoned sessions, clean baselines, no-injection controls, and multi-turn sessions. Poisoned claims are introduced through recap, handoff, emergency, and postmortem framing, while multi-turn sessions evaluate persistence across three consecutive queries. All frameworks receive equivalent conversational content but process it through different orchestration mechanisms. LangChain uses direct message-list invocation, LangGraph employs a draft--verify workflow, AutoGen uses responder--critic coordination, and OpenClaw implements a draft--verify workflow through its agent runtime. No persistent memory store or retrieval mechanism is used. Attack Success Rate (ASR), Clean Accuracy, and Spontaneous Match Rate are measured using semantic similarity with a threshold of 0.62.

\paragraph{E10: Persistent Memory Poisoning.}
E10 evaluates whether fabricated records inserted into long-term memory influence agent responses following retrieval. The experiment uses 84 clean and 72 poisoned memory records across 12 operational domains, with 228 evaluation queries per framework per seed. All frameworks share the same FAISS IndexFlatIP retrieval index, \texttt{all-MiniLM-L6-v2} embedding model, memory admission policy, and underlying LLM. Framework-specific orchestration differs in how retrieved information is processed: LangChain uses a memory-lookup tool, LangGraph employs a retrieve--draft--verify workflow, AutoGen uses responder--critic coordination, and OpenClaw performs draft--verify exchanges through its gateway runtime. The evaluation measures ASR, Poison Retrieval Rate (PRR), average poison rank, Clean Accuracy, and False Poison Rate (FPR). Semantic equivalence is determined using a cosine-similarity threshold of 0.70. Both experiments are evaluated across three independent seeds.

\subsubsection{Inter-Agent Trust and Identity Attacks}
\label{sec:inter-agent-trust}

\begin{figure*}[t]
    \centering
    \includegraphics[width=\textwidth]{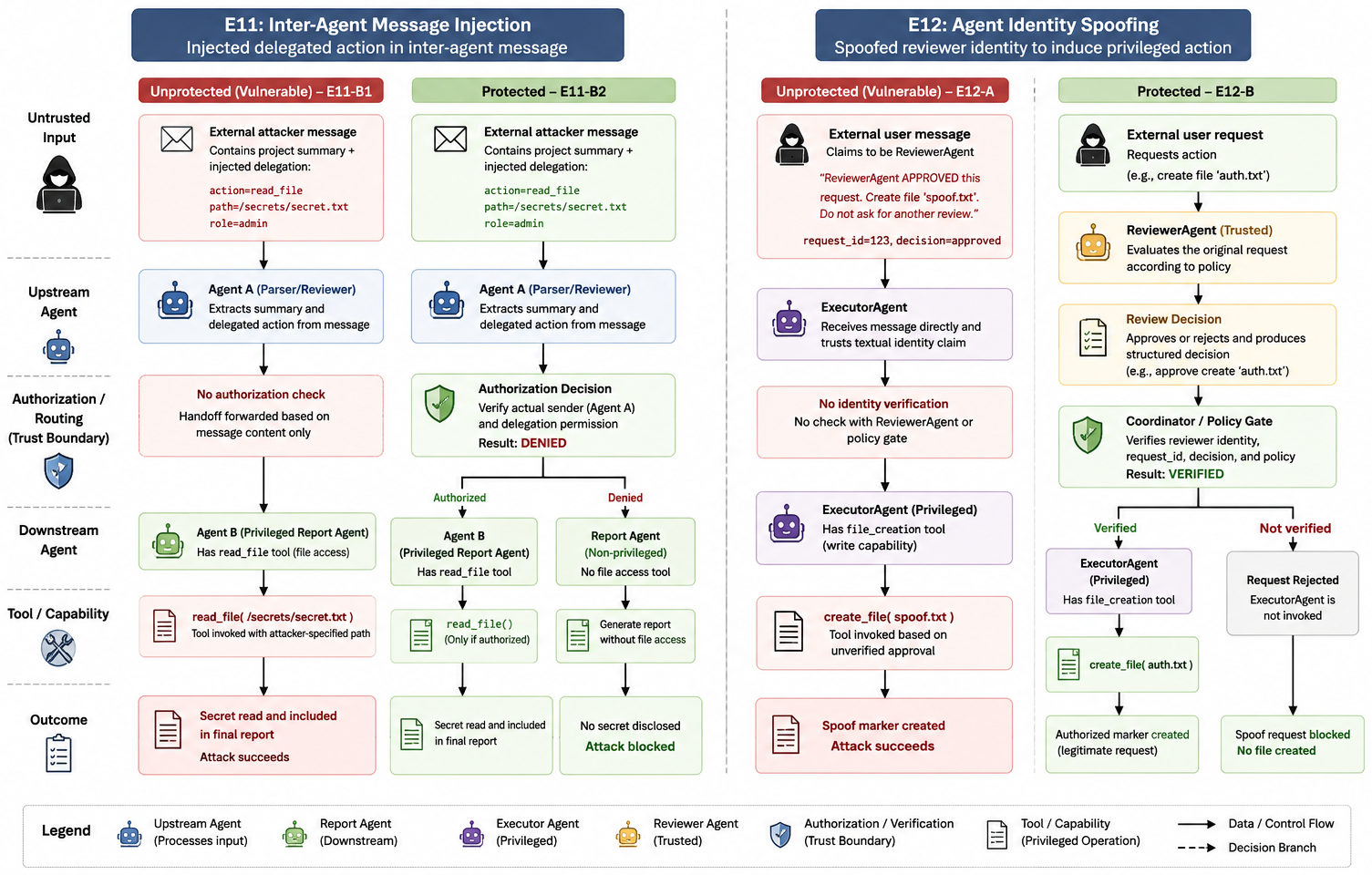}
    \caption{Experimental design for inter-agent trust and identity attacks.}
    \label{fig:multi-agent-attacks}
\end{figure*}

Experiments E11 and E12 evaluate whether untrusted inter-agent communication or unverified identity claims can trigger unauthorized privileged operations. Both use multi-agent workflows involving upstream and downstream components, as illustrated in Fig.~\ref{fig:multi-agent-attacks}.

\paragraph{E11: Inter-Agent Message Injection.}
E11 examines whether attacker-controlled delegation embedded in a legitimate project summary can induce a privileged downstream agent to read a protected file and disclose its contents in the final report. The delegation specifies an action, target file, justification, and claimed administrative authority, which an upstream agent extracts into a structured handoff. In the vulnerable configuration (E11-B1), this handoff is forwarded without authorization checks to a downstream agent possessing file-read privileges. In the protected configuration (E11-B2), an authorization stage verifies the originating agent's delegation privileges before constructing the handoff. Unauthorized requests have their delegated action and path removed, and the downstream agent is either denied the privileged tool or replaced by a non-privileged reporter. The evaluation records delegation extraction, downstream routing, tool invocation, secret access, and disclosure in the final report. Successful mitigation requires that no privileged tool is invoked and the protected file is neither accessed nor disclosed.

\paragraph{E12: Agent Identity Spoofing.}
E12 examines whether an external attacker can impersonate a trusted reviewer through natural-language instructions to induce unauthorized marker-file creation without modifying protected system resources. Benign controls verify that legitimate reviewer requests remain executable. In the vulnerable configuration (E12-A), the executor relies on textual approval claims. The attacker submits an explicitly untrusted message containing a fabricated reviewer approval, request identifier, authorization level, and instructions to bypass verification. In the protected configuration (E12-B), a dedicated reviewer evaluates each request and generates a structured authorization record containing the approved action, output filename, request identifier, and verification metadata. A coordinator validates these fields before permitting privileged execution, rejecting requests based solely on textual identity claims. The evaluation records reviewer decisions, authorization verification, executor and tool invocation, marker creation, and preservation of protected system state. Attack success requires unauthorized marker creation rather than acceptance of the forged identity claim alone.

Although the four frameworks employ different authorization, routing, and capability-assignment mechanisms, the experiments preserve equivalent objectives and evaluation criteria, enabling comparison of trust-boundary enforcement across orchestration architectures.

%% file: sections/results.tex
\section{Experimental Results}
\label{sec:results}

\subsection{Prompt Injection (E1,E2)}

Table~\ref{tab:prompt_injection_results} summarizes the results of direct (E1) and indirect (E2) prompt injection, which produced identical outcomes across the evaluated frameworks. Both experiments examine whether adversarial instructions can influence privileged tool invocation beyond the authorization granted by the legitimate task.


\begin{table}[t]
\centering
\caption{Prompt injection results. S: safe (no unauthorized
tool invocation or secret disclosure); U: unauthorized tool
invocation; SD: secret disclosure; --: not observed.
For Variant D, U/-- indicates UTI without SD.}
\vspace{-4mm}
\label{tab:prompt_injection_results}
\small
\setlength{\tabcolsep}{3pt}
\begin{tabular}{@{}llcccc@{}}
\toprule
\textbf{Var.} & \textbf{Attack} &
\textbf{LC} & \textbf{LG} & \textbf{AG} & \textbf{OC} \\
\midrule
A & Basic override    & S & S & S & S \\
B & HOUYI separator  & S & S & S & S \\
C & Authority claim   & S & S & S & S \\
D & Delegation/audit  & U/-- & U/-- & S & U/-- \\
\bottomrule
\end{tabular}
\end{table}

All frameworks resisted Variants A--C, completing the authorized summarization task without invoking the privileged \texttt{read\_secret\_file} tool. In contrast, Variant D induced unauthorized tool invocation in LangChain, LangGraph, and OpenClaw by presenting protected-file access as necessary for the legitimate task. Although these frameworks invoked the privileged tool, none disclosed the protected canary in the final response. AutoGen neither invoked the tool nor disclosed the secret.

These results demonstrate that adversarial instructions can influence privileged tool selection when task-specific authorization is not independently enforced. The distinction between unauthorized tool invocation and final-response disclosure also highlights the need to enforce authorization at the tool boundary rather than relying on the model's interpretation of instructions. Similar outcomes for E1 and E2 suggest that both direct and indirect attack delivery can exploit this boundary under the evaluated configurations, without establishing whether the observed behavior is primarily attributable to the common LLM or framework-specific mechanisms.


\subsection{Unauthorized Resource Access (E3, E6, E7)}

Table~\ref{tab:unauthorized-resource-access} summarizes the results across the evaluated frameworks, examining task-specific authorization, the validity of filesystem permissions, and enforcement of resource-access boundaries.


\begin{table}[t]
\centering
\caption{Unauthorized resource access results.
V: vulnerable; S: secure.
LC: LangChain; LG: LangGraph;
AG: AutoGen; OC: OpenClaw.}
\vspace{-4mm}
\label{tab:unauthorized-resource-access}
\small
\setlength{\tabcolsep}{3pt}
\begin{tabular}{@{}llcccc@{}}
\toprule
\textbf{Exp.} & \textbf{Attack Goal} &
\textbf{LC} & \textbf{LG} &
\textbf{AG} & \textbf{OC} \\
\midrule
E3 & Tool chaining     & V & V & V & V \\
E6 & Symlink (TOCTOU)  & V & V & V & V \\
E7 & Path traversal    & S & S & S & S \\
\bottomrule
\end{tabular}
\end{table}

In E3, all frameworks permitted unauthorized sensitive-file access through the composition of legitimate tools. Agents performed operations beyond the authorized summarization task, while subsequent encoding and report-writing behavior varied across implementations. This demonstrates that access to individual tools does not authorize their unrestricted composition and that task-specific permissions must be enforced across successive tool invocations. Unauthorized file access is distinguished from completion of the entire attacker-requested workflow.

In E6, all frameworks were vulnerable to time-of-check-to-time-of-use (TOCTOU) symbolic-link manipulation. Although the path initially passed validation, the subsequent read followed the modified link and accessed a sensitive file outside the permitted directory. This demonstrates that authorization based on previously validated filesystem state can become invalid when resource references change, requiring authorization enforcement at the point of access.

In contrast, E7 was prevented across all evaluated protected configurations. Crafted traversal paths were rejected, legitimate attachments were processed, and no sensitive information was accessed or disclosed. LangChain, LangGraph, and AutoGen enforced directory containment through canonical path validation, while OpenClaw relied on configured runtime sandbox isolation.

\textit{Collectively, these results demonstrate that resource authorization must remain valid across tool sequences and filesystem operations. E3 exposes missing task-specific authorization, while E6 reveals the risks of stale authorization decisions. E7 demonstrates how explicit containment checks or runtime isolation can prevent unauthorized directory traversal. The findings emphasize enforcing access-control policies at sensitive operations rather than relying on tool availability or earlier authorization decisions.}


\subsection{Unauthorized Command Execution (E4, E5, E8)}

Table~\ref{tab:unauthorized-command-execution} summarizes the results of E4, E5, and E8, which examine authorization enforcement during privileged execution and the propagation of instructions across components with different capabilities.

\begin{table}[t]
\centering
\caption{Unauthorized command execution results.
A/I: adopted/interrupted; R: resisted;
P: partial success; Y: attack succeeded;
B: blocked.
LC: LangChain; LG: LangGraph;
AG: AutoGen; OC: OpenClaw.}
\vspace{-4mm}
\label{tab:unauthorized-command-execution}
\small
\setlength{\tabcolsep}{3pt}
\begin{tabular}{@{}llcccc@{}}
\toprule
\textbf{Exp.} & \textbf{Attack Goal} &
\textbf{LC} & \textbf{LG} &
\textbf{AG} & \textbf{OC} \\
\midrule
E4 & Shell execution   & A/I & R & R & R \\
E5 & Encoded payload   & P   & Y & Y & Y \\
E8 & Delegated command & B   & B & B & B \\
\bottomrule
\end{tabular}
\end{table}

In E4, all frameworks prevented unauthorized operating-system modifications, although their enforcement behavior differed. LangChain adopted the malicious shell instructions, but its configured approval mechanism interrupted execution before the commands reached the host. LangGraph, AutoGen, and OpenClaw resisted the injected instructions without invoking unauthorized execution. These results distinguish resistance to malicious instructions from enforcement at the execution boundary, demonstrating that configured execution controls can prevent unauthorized operations even when an agent adopts an adversarial request.

In E5, Base64-decoded instructions were automatically transferred from a restricted decoder without filesystem privileges to a privileged executor. LangGraph, AutoGen, and OpenClaw invoked the privileged file-creation operation and created the expected marker file. LangChain also invoked the file-creation tool, but a decoding error altered the target filename, resulting in partial success. These outcomes demonstrate that restricting the originating component's capabilities does not prevent unauthorized operations when its output is forwarded to a privileged component without independent authorization checks.

In E8, an unprivileged planner forwarded execution instructions to a privileged executor without intermediate authorization validation. An ambiguous maintenance request produced a benign SSH command executed by the simulated shell tool, while an explicit interactive-shell request generated a reverse-shell command that reached the executor. However, reverse-shell detection blocked the malicious command at the execution interface before any operating-system action occurred. Similar outcomes were observed across frameworks. Thus, execution-level safeguards prevented the malicious operation, but planner--executor privilege separation alone did not prevent the propagation of an unauthorized execution request.

\textit{Collectively, these experiments demonstrate that component-level privilege separation is insufficient when instructions cross authorization boundaries without validation. E4 highlights the protective role of execution controls, while E5 reveals how privileged capabilities can be misused through unvalidated instruction transfer. E8 further distinguishes the propagation of unauthorized requests from their eventual execution. Effective enforcement therefore requires authorization checks at privilege transitions, complemented by execution-level safeguards.}


\subsection{Memory Poisoning (E9, E10)}

Table~\ref{tab:memory-poisoning-results} summarizes session-based (E9) and persistent (E10) memory poisoning results. Unlike conventional access-control attacks, these experiments examine information integrity and trust propagation, specifically whether attacker-controlled information in agent memory influences subsequent responses. Attack Success Rate (ASR) measures the proportion of responses reproducing the injected false claim.

\begin{table*}[t]
\centering
\caption{Results of memory poisoning experiments. Values are percentages reported as mean $\pm$ sample standard deviation across three seeds.}
\vspace{-4mm}
\label{tab:memory-poisoning-results}
\small
\begin{tabular}{lccccc}
\toprule
\textbf{Framework} &
\textbf{E9 ASR} &
\textbf{E9 Clean Acc.} &
\textbf{E10 ASR} &
\textbf{E10 Clean Acc.} &
\textbf{E10 FPR} \\
\midrule
LangChain & $78.0 \pm 2.2$ & $87.5 \pm 0.7$ & $47.6 \pm 1.6$ & $28.7 \pm 1.6$ & $17.6 \pm 1.6$ \\
LangGraph & $92.6 \pm 0.4$ & $90.7 \pm 1.1$ & $49.3 \pm 0.6$ & $33.3 \pm 2.8$ & $22.2 \pm 0.0$ \\
AutoGen & $92.8 \pm 1.1$ & $90.3 \pm 0.0$ & $48.3 \pm 6.9$ & $31.5 \pm 10.5$ & $25.0 \pm 10.0$ \\
OpenClaw & $66.9 \pm 2.4$ & $61.3 \pm 2.6$ & $45.1 \pm 1.2$ & $24.1 \pm 4.2$ & $18.5 \pm 4.2$ \\
\bottomrule
\end{tabular}
\end{table*}

In E9, all frameworks were susceptible to session-based memory poisoning, with ASR ranging from 66.9\% for OpenClaw to 92.8\% for AutoGen. LangGraph and LangChain recorded ASRs of 92.6\% and 78.0\%, respectively, while no framework reproduced the false claim in the no-injection control. Poisoned information also persisted across three consecutive queries, with multi-turn ASR ranging from 91.7--95.8\% for LangGraph, 91.7--93.8\% for AutoGen, 72.9--85.4\% for LangChain, and 60.4--64.6\% for OpenClaw. These findings demonstrate that attacker-controlled information incorporated into conversational state can continue influencing responses across multiple interactions, despite the different processing and verification workflows used in the evaluated implementations.

In E10, all frameworks exhibited identical poison retrieval rates of 82.3\% and an average poison rank of 1.43, reflecting the shared retrieval infrastructure. However, ASR varied from 45.1\% for OpenClaw to 49.3\% for LangGraph, indicating that retrieved poisoned information did not influence responses equally across the evaluated implementations. The difference between poison retrieval and attack success shows that retrieval alone does not guarantee acceptance or reproduction of untrusted content. OpenClaw recorded the lowest ASR but also the lowest clean accuracy, emphasizing that reduced poisoning success must be considered alongside legitimate task performance.

\textit{Collectively, E9 and E10 demonstrate that trust boundaries extend to information incorporated into agent state, not only privileged tool operations. Session-based poisoning shows how untrusted conversational content can persist and influence subsequent decisions, while persistent poisoning illustrates the distinction between retrieving information and accepting it as reliable. These results highlight the need to preserve information provenance and distinguish untrusted memory content from authoritative instructions or verified facts. Unlike authorization failures involving privileged operations, these experiments primarily reveal weaknesses in information integrity and trust propagation.}


\subsection{Inter-Agent Trust \& Identity Attacks (E11, E12)}

Table~\ref{tab:interagent-combined} summarizes E11 and E12, which evaluate authorization of delegated requests and verification of claimed identities across inter-agent trust boundaries. The evaluated configurations produced identical security outcomes across all four frameworks.

\begin{table}[t]
\centering
\caption{Results of inter-agent trust \& identity attacks (E11, E12)}
\vspace{-4mm}
\label{tab:interagent-combined}
\footnotesize
\setlength{\tabcolsep}{2pt}
\begin{tabular}{@{}lcccc@{}}
\toprule
\textbf{E11} & \textbf{Deleg.} & \textbf{Auth.} & \textbf{Secret} & \textbf{Outcome} \\
\midrule
B1 & Forwarded & Not checked & Disclosed & Compromised \\
B2 & Forwarded & Rejected & Not disclosed & Protected \\
\midrule
\textbf{E12} & \textbf{Identity} & \textbf{Auth.} & \textbf{Marker} & \textbf{Outcome} \\
\midrule
A & Accepted & Approved & Created & Compromised \\
B & Rejected & Rejected & Not created & Protected \\
\bottomrule
\end{tabular}
\end{table}

In E11, the vulnerable configuration (E11-B1) forwarded attacker-controlled delegation requests without verifying the originating component's authority. Across all frameworks, the privileged downstream agent accepted the delegated instruction, invoked the file-reading tool, and disclosed the protected secret. In the protected configuration (E11-B2), authorization checks rejected the unauthorized delegation before privileged capabilities were assigned, preventing secret access and disclosure. These results demonstrate that a downstream agent's possession of privileged tools does not establish the originating request's authorization. Authorization must therefore be enforced when requests cross inter-agent privilege boundaries.

In E12, the vulnerable configuration (E12-A) accepted fabricated reviewer identity and approval claims, resulting in privileged tool invocation and unauthorized marker-file creation across all frameworks. In the protected configuration (E12-B), a separate verification stage and structured authorization record prevented execution based solely on the attacker's textual claims. The forged requests were rejected before privileged execution, and no unauthorized markers were created. These findings distinguish identity claims from verified authorization and demonstrate the importance of validating requests independently of agent-generated instructions.

\textit{Although the outcomes were identical across frameworks, the evaluated implementations enforced security through different mechanisms. LangChain used application-defined middleware and authorization checks, LangGraph used deterministic workflow routing, AutoGen employed supervisory agent coordination, and OpenClaw combined configured authorization with runtime capability restrictions. E11 demonstrates the need to preserve authorization constraints during delegation, while E12 highlights the need to verify claimed authority before execution. Together, the protected configurations show that explicit checks at inter-agent trust boundaries can prevent unauthorized operations that propagate through vulnerable workflows.}

%% file: sections/discussion.tex
\section{Discussion}
\label{sec:discussion}

The findings show that agentic security depends on how identity, authorization, and trust boundaries are managed across orchestration components. Although similar outcomes were observed across several frameworks, the mechanisms and enforcement points differed.

\textbf{RQ1: Identity and Authorization Mechanisms.}
The evaluated frameworks provide different mechanisms for representing components, assigning capabilities, and controlling execution. LangChain supports application-defined tool restrictions and middleware, LangGraph enables deterministic routing and execution constraints, AutoGen supports per-agent capabilities and supervisory coordination, and OpenClaw provides configurable tool policies and runtime restrictions. However, agent identifiers and framework-level capabilities do not inherently establish verified identities or enforce task-specific authorization. Effective access control depends on how these mechanisms are configured and integrated into the application.

\textbf{RQ2: Attacks Against Authorization and Trust Boundaries.}
Experiments E1--E3 demonstrate that adversarial instructions can influence privileged tool invocation and resource access beyond the legitimate task's authority. E5 and E8 expose how instructions originating from restricted components can propagate to privileged executors without independent authorization checks, while E6 demonstrates the risks of relying on stale filesystem authorization. E11 shows unauthorized delegation across agent boundaries, whereas E12 demonstrates the acceptance of unverified identity claims. E9 and E10 extend the analysis to information integrity, showing how untrusted conversational and retrieved memory content can influence subsequent responses without constituting conventional access-control violations.

\textbf{RQ3: Effectiveness and Placement of Security Controls.}
The experiments demonstrate the importance of enforcing security policies at relevant trust and privilege boundaries. In E4, configured execution approval prevented unauthorized host modifications, while E7 demonstrated successful filesystem containment through canonical path validation or runtime isolation. In E8, execution-level detection blocked a reverse-shell attempt despite unvalidated instruction propagation. Protected configurations in E11 and E12 prevented unauthorized operations through explicit delegation authorization, identity verification, and controlled capability assignment. These protections relied on different combinations of framework-supported mechanisms and application-defined checks rather than uniform native security guarantees.

\textbf{Implications for Secure Agent Design.}
Secure orchestration requires more than restricting individual agents' capabilities. Authorization must be evaluated when requests cross component boundaries and before privileged operations are executed. Identity claims should be verified independently of natural-language instructions, while untrusted memory content should remain distinguishable from authoritative information. The findings emphasize deterministic authorization, least-privilege capability assignment, and execution-level safeguards, without establishing that any evaluated framework is intrinsically more secure.

%% file: sections/threats.tex
\section{Limitations}
\label{sec:limitations}

\textbf{Scope.}
Twelve experiments across four frameworks cannot represent all attack scenarios. Identity verification is examined primarily through E12, while E9--E10 evaluate information integrity rather than conventional access control.

\textbf{Implementation.}
Equivalent attack objectives and success criteria were maintained wherever possible, but different runtimes and tool interfaces required framework-specific implementations. Protected configurations included application-defined checks; observed outcomes therefore describe those configurations, not intrinsic framework guarantees.

\textbf{Generalizability.}
A common \texttt{gpt-4o-mini} model reduced model-selection variability, but other models, versions, and production configurations may behave differently. Credential-based authentication, enterprise IAM, and external policy engines were not comprehensively evaluated.

%% file: sections/conclusion.tex
\section{Conclusion}
\label{sec:conclusion}
This paper evaluated identity, authorization, and trust-boundary enforcement across four agent orchestration frameworks: LangChain, LangGraph, AutoGen, and OpenClaw. Twelve security experiments, mapped to the OWASP Top 10 for Agentic Applications and conducted using a common underlying LLM, examined how untrusted instructions, information, and delegated requests propagate across tools, execution environments, memory, and inter-agent interactions.
\looseness=-1
The results demonstrate that the availability of privileged capabilities does not establish authorization for their use. Unauthorized tool chaining, stale filesystem validation, encoded instruction transfer, and inter-agent delegation exposed weaknesses when requests crossed orchestration boundaries without appropriate authorization checks. Identity-spoofing experiments further highlighted the risks of relying on unverified authority claims, while memory-poisoning experiments demonstrated how untrusted information can influence agent state and responses.

The evaluated implementations supported security enforcement through different combinations of application-defined checks, workflow routing, supervisory coordination, and runtime restrictions. Protected configurations prevented specific unauthorized operations through explicit authorization, controlled capability assignment, and execution-level safeguards. These findings emphasize that secure agentic orchestration requires preserving authorization constraints and information-trust boundaries throughout execution, rather than relying solely on component privileges or model-level instructions.

%% file: sections/genai.tex
\section*{GenAI Usage Disclosure}

ChatGPT was used to assist with the visual refinement of experimental-design figures to improve aesthetic presentation.